\documentclass[11pt]{article}
\usepackage{amsmath,amssymb}
\usepackage[dvips]{epsfig}
\usepackage{graphicx}

{\hspace*{\fill}$\rule{.3\baselineskip}{.35\baselineskip}$\end{trivlist}}

\renewcommand{\geq}{\geqslant}
\renewcommand{\leq}{\leqslant}

\renewcommand{\phi}{\varphi}
\newcommand{\be}{\begin{eqnarray}}
\newcommand{\ee}{\end{eqnarray}}

\newcommand{\eps}{\varepsilon}

\begin{document}

\title{\bf On the characterization of vortex configurations\\
in the steady rotating Bose--Einstein condensates}

\author{P.G. Kevrekidis$^1$ and D.E. Pelinovsky$^2$ \\
{\small $^{1}$ Department of Mathematics and Statistics, University
of Massachusetts, Amherst, MA 01003} \\
{\small $^{2}$ Department of Mathematics and Statistics, McMaster
University, Hamilton, Ontario, Canada, L8S 4K1} }

\date{\today}
\maketitle

\begin{abstract}
Motivated by experiments in atomic Bose-Einstein condensates (BECs), we compare predictions of a
system of ordinary differential equations (ODE) for dynamics of one and two individual vortices
in the rotating BECs with those of the partial differential equation (PDE).
In particular, we characterize orbitally stable vortex configurations in a symmetric harmonic trap
due to a cubic repulsive interaction and the steady rotation. The ODE system is analyzed in details
and the PDE model is approximated numerically. Good agreement between the two models is established
in the semi-classical (Thomas-Fermi) limit that corresponds to the BECs at the large chemical potentials.
\end{abstract}



\section{Introduction}

Our principal interest in the present work focuses on the dynamics of vortex excitations in atomic Bose-Einstein
condensates~\cite{pitas} and their description with the Gross--Pitaevskii (GP) equation~\cite{siambook}.
Early work on the subject, summarized in the review \cite{review}, as well
as more recent experimental work such as in \cite{Navarro}
highlight the ongoing interest towards a quantitative characterization
of vortex configurations of minimal energy by means of low-dimensional models
involving ordinary differential equations (ODEs). This is an endeavor that was initiated
in the pioneering work of~\cite{Castin} and has now matured to the
point that it can be used to understand the dynamics of such systems
in experimental time series such as those of~\cite{Navarro}
(see also the relevant analysis of~\cite{Zamp}).
Our aim in the present work is to characterize orbitally stable vortex configurations
among steadily rotating solutions to the GP equation.

More specifically, we address the GP equation for a Bose--Einstein condensate (BEC)
in two dimensions with a cubic repulsive interaction and a
symmetric harmonic trap. This model can be written in the normalized form
\begin{equation}
i \eps u_{t} = -\eps^2 \Delta u + (|x|^2 + |u|^2 -1) u, \label{GP}%
\end{equation}
where $\Delta = \partial_x^2 + \partial_y^2$ and $|x|^2 = x^2 + y^2$.
By means of the transformation $u = \sqrt{\eps} \tilde{u}$
and $x = \sqrt{\eps} \tilde{x}$, the model can be rewritten in the form
\begin{equation}
i \tilde{u}_{t} = -\tilde{\Delta} \tilde{u} + (|\tilde{x}|^2 + |\tilde{u}|^2 - \mu) \tilde{u}, \label{GPtilde}%
\end{equation}
where $\mu = \eps^{-1}$ is the chemical potential.
Naturally, the regime where $\eps$ is a small parameter
corresponds to the regime of the large chemical potential $\mu$.
In this semi-classical (Thomas--Fermi) limit $\eps \to 0$,
vortices behave qualitatively as individual particles with
no internal structure~\cite{siambook}.

The associated energy of the GP equation (\ref{GP}) is given by
\begin{equation}
E(u) = \int\!\int_{\mathbb{R}^{2}} \left[  \eps^2 |\nabla u|^{2} + (|x|^{2}-1) |u|^{2} + \frac{1}{2} |u|^{4}\right]  dx dy. \label{Energy}%
\end{equation}
Time-independent solutions to the GP equation (\ref{GP}) are critical points of the energy (\ref{Energy}).

Among the stationary solutions of the GP equation (\ref{GP}), there is a {\em ground state} (global minimizer)
of energy $E(u)$ subject to a positive value of mass $Q(u) = \| u \|^2$. The ground state
is a radially symmetric, real, positive stationary solution with a fast decay to zero at infinity.
Properties of the ground state in the semi-classical limit $\eps \to 0$ were
studied in \cite{GalloPel,IM1}. On the other hand, vortices are complex-valued stationary solutions
with a nonzero winding number along a circle of large radius centered at the origin.
Vortices are less energetically favorable, as they are saddle points of energy $E(u)$
subject to the positive value of mass $Q(u)$.
However, when the BEC is rotated with a constant angular frequency $\omega$,
it was realized long ago~\cite{review}
that the vortex configurations may become energetically favorable
depending on the frequency $\omega$ due to the contribution
of the $z$-component of the angular momentum in the total energy.

From a mathematical perspective, Ignat and Millot \cite{IM1,IM2} confirmed that
the vortex of charge one near the center of symmetry is a global minimizer of total energy
for a frequency $\omega$ above a first critical value $\omega_1^*$. Seiringer \cite{Seiringer}
proved that a vortex configuration with charge $m$ becomes energetically favorable to a vortex configuration
with charge $(m-1)$ for a  frequency $\omega$ above the $m$-th critical value $\omega_m^* > \omega_{m-1}^*$
and that radially symmetric vortices with charge $m \geq2$ cannot be
minimizers of total energy. It is natural to conjecture that the vortex configuration
of charge $m$ with the minimal total energy consists of $m$ individual vortices of charge
one, which are placed near the center of symmetry. The location of individual vortices has not been rigorously
discussed in the previous works \cite{IM1,IM2,Seiringer}, although it has
been the subject of studies in the physical literature (see relevant examples in \cite{Castin,Navarro,Zamp}).

For the vortex of charge one, it was shown by using variational approximations
\cite{Castin} and bifurcation methods \cite{PeKe13} that the radially symmetric vortex
becomes a local minimizer of total energy past the threshold value $\omega_1$
of the rotation frequency $\omega$, where $\omega_1 \leq \omega_1^*$.
In addition to the radially symmetric vortex, which exists for all values of $\omega$,
there exists another branch of the asymmetric vortex solutions above the threshold value,
for $\omega > \omega_1$. The branch is represented by a vortex of charge one
displaced from the center of rotating symmetric trap. Although the asymmetric vortex
is not a local energy minimizer, it is nevertheless a constrained energy minimizer
subject to the constraint eliminating the rotational invariance of the asymmetric vortex.
Consequently, both radially symmetric and asymmetric
vortices are orbitally stable in the time evolution of the
GP equation (\ref{GP}) for the rotation frequency $\omega$ slightly above the
threshold value $\omega_1$ \cite{PeKe13}.

Stability of equilibrium configurations of several vortices of charge one in rotating harmonic
traps was investigated numerically in \cite{Kollar,21,34,38,PeKe11,57}
(although a number of these studies have involved also vortices of opposite
charge). The numerical results were compared with the predictions given by the finite-dimensional system
for dynamics of individual vortices \cite{theo2,theo1,Navarro,Zamp}.
The relevant dynamics even for systems of two vortices remain a topic
of active theoretical investigation~\cite{simula}.

In the case of two vortices, the equilibrium configuration of minimal total energy
emerges again above the threshold value $\omega_2$ for the rotation frequency $\omega$,
where $\omega_2 > \omega_1$. The relevant configuration consists of two vortices of charge one
being located symmetrically with respect to the center of the harmonic trap. However,
the symmetric vortex pair is stable only for small distances from the center and
it loses stability for larger distances  \cite{Navarro}. Once it becomes unstable, another asymmetric configuration
involving two vortices bifurcates with one vortex being at a smaller-than-critical distance from
the center and the other vortex being at a larger-than-critical distance from the
center. The asymmetric pair is stable in numerical simulations and coexists
for rotating frequencies above the threshold value with the stable symmetric
vortex pair located at the smaller-than-critical distances~\cite{Navarro,Zamp}.

In this work, we revisit the ODE models for configurations of two vortices of charge one
in the semi-classical limit $\eps \to 0$. We will connect the details of bifurcations observed
in \cite{Navarro,Zamp} with the stability properties of vortices due to their energy minimization properties.
Compared to our previous work \cite{PeKe11},
we will incorporate an additional term in expansion of vortex's kinetic energy,
which is responsible for the nonlinear dependence of the vortex precession frequency on the
vortex distance from the origin. This improvement corresponds exactly
to the theory used in the physics literature; see, e.g., the review \cite{review}.
The additional term in the total energy derived in Appendix A allows us to give all
details on the characterization of energy minimizers and orbital stability in the case
of one and two vortices of charge one.

In particular, we recover the conclusions obtained from the bifurcation theory
in~\cite{PeKe13} that the symmetric vortex of charge one is an energy minimizer for $\omega > \omega_1$
and that the asymmetric vortex of charge one is a constrained energy minimizer for $\omega > \omega_1$.
Both vortex configurations are stable in the time evolution of the GP equation (\ref{GP}).

We also show from the ODE model that the symmetric pair of two vortices of charge one
is an energy minimizer for $\omega > \omega_2$, whereas the asymmetric pair is
a local constrained minimizer of energy
for $\omega > \omega_2$. In this case too, for $\omega > \omega_2$,
both vortex configurations are stable in the time evolution
of the GP equation (\ref{GP}). A fold bifurcation
of the symmetric vortex pair occurs at a frequency $\omega$ smaller than $\omega_2$
with both branches of symmetric vortex pairs being unstable near the fold bifurcation.
This instability is due to the symmetric vortex pairs for $\omega < \omega_2$
being saddle points of total energy even in the presence of the constraint eliminating
rotational invariance of the vortex configuration.

Although the ODE model is not rigorously justified in the context of the GP equation (\ref{GP}),
we confirm numerically that the predictions of the ODE model hold
exactly as qualitatively predicted within the PDE model
in the semi-classical limit $\eps \to 0$.

Next, we mention a number of recent studies
on vortex configurations of the GP equation (\ref{GP}) in the case of steady rotation.
In the small-amplitude limit, when the reduced models are derived by
using the decompositions over the Hermite--Gauss eigenfunctions of the quantum harmonic oscillator,
classification of localized (soliton and vortex) solutions from the triple
eigenvalue was constructed in \cite{Kap1}. Bifurcations of
radially symmetric vortices with charge $m\in\mathbb{N}$ and dipole solutions
were studied in \cite{CoGa15} with the help of the equivariant degree theory.
Bifurcations of multi-vortex configurations in the parameter continuation
with respect to the rotation frequency $\omega$ were considered in \cite{GarciaPel}.
Existence and stability of stationary states were analyzed in \cite{Hani} with the
resonant normal forms. Some exact solutions of the resonant normal forms
were reported recently in \cite{Bizon}. Vortex dipoles were studied with the normal form
equations in the presence of an anisotropic trap in~\cite{GKC}.

Compared to the recent works developed in the small-amplitude limit, our results here are formally valid only
in the semi-classical limit $\eps \to 0$, i.e., for large
chemical potential $\mu$ rather than for values of the chemical
potential in the vicinity of the linear limit.
As a result, our conclusions are slightly different
from those that hold in the small-amplitude limit.

In \cite{GarciaPel}, it was shown that the asymmetric pair of two vortices
of charge one bifurcates from the symmetric vortex of charge $m = 2$
and that this vortex pair shares the instability of the symmetric vortex of charge $m = 2$
in the small-amplitude limit. This instability is due to the fact that
the vortex pair is a saddle point of total energy above the bifurcation threshold.
It is presently an open question to explore
how this bifurcation diagram deforms when the chemical potential is
changed from the small-amplitude
limit to the semi-classical (Thomas--Fermi) limit.

Recent computational explorations of the stationary configurations of vortices have been
performed with several alternative numerical methods \cite{Farell,Bartek,theo}.
A principal direction of attention
is drawn to the global minimizers of total energy in the case of fast rotation, when the computational
domain is filled with the triangular lattice of vortices~\cite{Bartek,theo}. Dissipation is also included
in order to regularize the computational algorithms \cite{theo} or to enable
convergence in the case of ground states \cite{Bartek}. Although
the ODE models are very useful to characterize one and two vortices, it becomes cumbersome
to characterize three and more vortices, and naturally the complexity increases significantly in
the case of larger clusters and especially for triangular vortex lattices.
Hence, such cases will not be addressed, although the tools utilized here
can in principle be generalized therein.

Our work paves the way for numerous developments in the future. Constructing
multi-vortex configurations and lattices of such vortices in a systematic way at the ODE level
is definitely a challenging problem for better understanding of dynamics in the GP equation.
Another important direction of recent explorations in BECs has involved the
phenomenology of vortex lines and vortex rings in the space of three dimensions~\cite{siambook}.
The consideration of similar notions of effective dynamical systems
describing, e.g., multiple vortex rings is a topic under active
investigation and one that bears some nontrivial challenges from
the ODE theory~\cite{bisset}.

Finally, we mention that vortex ODE theory has been found very useful
to characterize travelling waves in the defocusing nonlinear Schr\"{o}dinger
equation in the absence of rotation and the harmonic potential \cite{Bet1,Bet2}
(see also the recent work \cite{Chiron}).

The remainder of this paper is organized as follows. Section 2 reports
predictions of the ODE
model for a single vortex of charge one. Section 3 is devoted to analysis of
the ODE model for
a pair of vortices of charge one. Section 4 gives numerical results for the
vortex pairs.
Section 5 presents our conclusions. Appendix A contains
derivation of the additional term in expansion of vortex's kinetic energy.

\section{Reduced energy for a single vortex of charge one}

A single vortex of charge one shifted from the center of the harmonic potential behaves like a particle with
the corresponding kinetic and potential energy \cite{siambook}. The asymptotic expansions
of vortex's kinetic and potential energy were derived in \cite{PeKe11} by using a formal Rayleigh--Ritz method
and analysis of resulting integrals in the semiclassical limit of $\eps \to 0$.
By Lemmas 1 and 2 in \cite{PeKe11}, a single vortex of charge one located at the point
$(x_0,y_0) \in \mathbb{R}^2$ has kinetic $K$ and potential $P$ energies given by
\begin{equation}
\label{kinetic-1}
K(x_0,y_0) = \frac{1}{2} \eps (x_0 \dot{y}_0 - y_0 \dot{x}_0) \left[ 1 + \mathcal{O}(\eps + x_0^2 + y_0^2) \right]
\end{equation}
and
\begin{equation}
\label{potential-1}
P(x_0,y_0) =  \frac{1}{2} \eps \omega_0(\eps) (x_0^2+y_0^2) \left[ 1 + \mathcal{O}(\eps^{1/3} + x_0^2+y_0^2) \right],
\end{equation}
where $\omega_0(\eps) = -2 \eps \log(\eps) + \mathcal{O}(1)$ as $\eps \to 0$ and we have divided all expressions by
$2\pi$ compared to \cite{PeKe11}. Let us truncate the expansions (\ref{kinetic-1}) and (\ref{potential-1})
by the leading-order terms
and obtain the Euler--Lagrange equations for the Lagrangian $L(x_0,y_0) = K(x_0,y_0) - P(x_0,y_0)$.
The corresponding linear system divided by $\eps$ is
\begin{equation}
\label{dynamics-1}
\left\{ \begin{array}{l} \dot{y}_0 - \omega_0(\eps) x_0 = 0, \\
- \dot{x}_0 - \omega_0(\eps) y_0 = 0, \end{array} \right. \quad \Rightarrow
\quad \ddot{x}_0 + \omega_0(\eps)^2 x_0 = 0,
\end{equation}
and it exhibits harmonic oscillators with the frequency $\omega_0(\eps)$. This frequency
was compared in \cite{PeKe11} with the smallest eigenvalue in the spectral stability problem
for the single vortex of charge one obtained numerically,
a good agreement was found in the asymptotic limit $\eps \to 0$.

It was suggested heuristically in \cite{review} (see also \cite{theo2,theo1})
that the frequency of vortex precession depends on the displacement $a$ from the center
of the harmonic potential by the following law
\begin{equation}
\label{omega-law}
\omega(a) = \frac{\omega_0(\eps)}{1 - a^2}, \quad a \in (0,1),
\end{equation}
so that $\omega(a) > \omega_0(\eps)$. This law is in agreement with the bifurcation
theory for a single asymmetric vortex in the stationary GP equation \cite{PeKe13},
where a new branch of stationary vortex solutions displaced from
the center of the harmonic potential by the distance $a \sim (\omega - \omega_0(\eps))^{1/2}$
was shown to exist for $\omega \gtrsim \omega_0(\eps)$.

The empirical law (\ref{omega-law}) and the bifurcation of asymmetric vortices for $\omega \gtrsim \omega_0(\eps)$
can be explained by the extension of the kinetic energy given by (\ref{kinetic-1}) at the same order of $\eps$
but to the higher order in $x_0^2+y_0^2$. We show in Appendix A that
the kinetic energy $K(x_0,y_0)$ can be further expanded as follows:
\begin{equation}
\label{kinetic-2}
K(x_0,y_0) = \frac{1}{2} \eps (x_0 \dot{y}_0 - y_0 \dot{x}_0) \left[ 1 - \frac{1}{2} (x_0^2 + y_0^2)
+ \mathcal{O}\left(\eps + x_0^4 + y_0^4 \right) \right].
\end{equation}
In the reference frame rotating with the angular frequency $\omega$, we can use the polar coordinates
\begin{equation}
\label{rotating-frame}
x_0 = \xi_0 \cos(\omega t) - \eta_0 \sin(\omega t), \quad y_0 = \xi_0 \sin(\omega t) + \eta_0 \cos(\omega t)
\end{equation}
and rewrite the truncated kinetic and potential energies as follows:
\begin{eqnarray*}
K(\xi_0,\eta_0) & = & \frac{1}{2} \eps (\xi_0 \dot{\eta}_0 - \dot{\xi}_0 \eta_0)
+ \frac{1}{2} \eps \omega (\xi_0^2+\eta_0^2)  \left[ 1 - \frac{1}{2} (\xi_0^2+\eta_0^2) \right], \\
P(\xi_0,\eta_0) & = & \frac{1}{2} \eps \omega_0(\eps) (\xi_0^2+\eta_0^2),
\end{eqnarray*}
where the nonlinear correction in front of $(\xi_0 \dot{\eta}_0 - \dot{\xi}_0 \eta_0)$ in $K(\xi_0,\eta_0)$
is dropped to simplify the time evolution of the ODE system.
In the remainder of this section, we review the existence and stability results
for the single vortex of charge one within the ODE theory.

\subsection{Existence of steadily rotating vortices}

Steadily rotating vortices are critical points of the action functional
\begin{eqnarray}
\label{reduced-1}
E_1(\xi_0,\eta_0) = \frac{1}{2} \eps \omega (\xi_0^2 + \eta_0^2) \left[ 1 - \frac{1}{2} (\xi_0^2+\eta_0^2) \right]
- \frac{1}{2} \eps \omega_0(\eps) (\xi_0^2+\eta_0^2).
\end{eqnarray}
Thanks to the rotational invariance, one can place the steadily rotating vortex to the point
$(\xi_0,\eta_0) = (a,0)$. The Euler--Lagrange equation for $E_1(a,0)$ yields
$$
\frac{d}{da} E_1(a,0) = \eps \omega a (1 - a^2) - \eps \omega_0(\eps) a = 0.
$$
Two solutions exists: one with $a = 0$ for every $\omega$ and the other one with $a \in (0,1)$
for $\omega(a)$ given by the dependence (\ref{omega-law}).
The symmetric vortex with $a = 0$ exists for every $\omega$, whereas
the asymmetric vortex with the displacement $a > 0$ exists for $\omega \gtrsim \omega_0(\eps)$.

\subsection{Variational characterization of the individual vortices}

Extremal properties of the two critical points of $E_1(\xi_0,\eta_0)$ are studied from
the Hessian matrix $E_1''(a,0)$. This is a diagonal matrix with the diagonal entries:
$$
\partial_{\xi_0}^2 E_1(a,0) = \eps \omega (1-3a^2) - \eps \omega_0(\eps),
\quad \partial_{\eta_0}^2 E_1(a,0) = \eps \omega (1-a^2) - \eps \omega_0(\eps).
$$
The critical point $(0,0)$ is a minimum of $E_1$ for $\omega > \omega_0(\eps)$
and a saddle point of $E_1$ with two negative eigenvalues if $\omega < \omega_0(\eps)$.
The critical point $(a,0)$ with $a > 0$ and $\omega > \omega_0(\eps)$ related by
equation (\ref{omega-law})  is a saddle point of $E_1$ with one negative and one zero eigenvalues.
This conclusion agrees with the full bifurcation analysis of the GP equation (\ref{GP}) given in \cite{GarciaPel,PeKe13}.

The zero eigenvalue for the asymmetric vortex with $a > 0$ is related to the rotational invariance
of the vortex configuration, which can be placed at any $(\xi_0,\eta_0) = a (\cos \alpha, \sin \alpha)$
with arbitrary $\alpha \in [0,2\pi]$. The corresponding eigenvector in the kernel
of $E_1''(a,0)$ is $R := (0,1)^T$.

\subsection{Stability of steadily rotating vortices}

Stability of the two critical points of $E_1(\xi_0,\eta_0)$ is determined by
equations of motion obtained from the leading-order Lagrangian
$$
L_1(\xi_0,\eta_0) = \frac{1}{2} \eps (\xi_0 \dot{\eta}_0 - \dot{\xi}_0 \eta_0) + E_1(\xi_0,\eta_0).
$$
After dividing the Euler--Lagrange equations by $\eps$, equations of motion take the form
\begin{eqnarray*}
\left\{ \begin{array}{l} \dot{\eta}_0 + \omega \xi_0 ( 1 - \xi_0^2 - \eta_0^2) - \omega_0(\eps) \xi_0 = 0,\\
\dot{\xi}_0 - \omega \eta_0 ( 1 - \xi_0^2 - \eta_0^2) + \omega_0(\eps) \eta_0 = 0, \end{array} \right.
\end{eqnarray*}
which can be written as the Hamiltonian system
\begin{equation}
\label{ham-sys-1}
\frac{d}{d t} \left( \begin{array}{c} \xi_0 \\ \eta_0 \end{array} \right) =
J \left( \begin{array}{c} \frac{\partial E_1}{\partial \xi_0} \\
\frac{\partial E_1}{\partial \eta_0} \end{array} \right),
\quad J = \left[ \begin{array}{cc} 0 & 1 \\ -1 & 0 \end{array} \right],
\end{equation}
where $E_1$ in (\ref{reduced-1}) serves as the Hamiltonian function.

Spectral stability of the two vortex solutions can be analyzed from the linearization of the
Hamiltonian system (\ref{ham-sys-1}) at the critical point $(\xi_0,\eta_0) = (a,0)$.
Substituting $\xi_0 = a + \hat{\xi}_0 e^{\lambda t}$, $\eta_0 = \hat{\eta}_0 e^{\lambda t}$
and neglecting the quadratic terms in $(\hat{\xi}_0,\hat{\eta}_0)$ yield the spectral stability problem
\begin{equation}
\label{spectral-stab-1}
\left\{ \begin{array}{l} \left[ \omega (1-3a^2) - \omega_0(\eps) \right] \hat{\xi}_0 = - \lambda \hat{\eta}_0, \\
\left[ \omega (1-a^2) - \omega_0(\eps) \right] \hat{\eta}_0 = \lambda \hat{\xi}_0. \end{array} \right.
\end{equation}

For the symmetric vortex with $a = 0$, the spectral problem (\ref{spectral-stab-1}) admits a pair of purely
imaginary eigenvalues with
$$
\lambda^2 = -(\omega - \omega_0(\eps))^2,
$$
both for $\omega < \omega_0(\eps)$ and $\omega > \omega_0(\eps)$.
For the asymmetric vortex with $a > 0$ and $\omega > \omega_0(\eps)$ related by equation (\ref{omega-law}),
the spectral problem (\ref{spectral-stab-1}) admits a double zero eigenvalue.
These conclusions of the ODE theory agree with the numerical results obtained for the PDE model (\ref{GP}) in \cite{PeKe13}.
In particular, both the symmetric and asymmetric vortices were found to be spectrally stable for $\omega$ near $\omega_0(\eps)$.
The symmetric vortex was found to have a pair of purely imaginary eigenvalues near the origin
coalescing at the origin for $\omega = \omega_0(\eps)$. The asymmetric vortex was found to have an additional
degeneracy of the zero eigenvalue due to the rotational symmetry.

The spectral (and orbital) stability of the asymmetric vortex is explained by its energetic
characterization. While the critical point $(a,0)$ is a saddle point of $E_1$,
it is a constrained minimizer of $E_1$ under the constraint eliminating the rotational symmetry
and preserving the symplectic structure of the Hamiltonian system (\ref{ham-sys-1}).
Since $R = (0,1)^T$ spans the kernel of the Hessian matrix $E_1''(a,0)$, the symplectic orthogonality constraint takes the form
\begin{equation}
\label{constraint-1}
\phi := (\xi_0,\eta_0)^T \in \mathbb{R}^2 : \quad \langle J^{-1} \phi, R \rangle  = 0,
\end{equation}
which simplifies to $\xi_0 = 0$. The constraint $\xi_0 = 0$ removes the negative eigenvalue
of the Hessian matrix $E_1''(a,0)$. Hence, the critical point $(a,0)$
is a constrained minimizer of $E_1$ under the constraint (\ref{constraint-1})
related to the rotational invariance.

\section{Reduced energy for a pair of vortices of charge one}

We now turn to the examination of a pair of vortices of charge one.
It was argued in \cite{theo2,theo1} that dynamics of two and more individual vortices can be modeled
by using the reduced energy, which is given by the sum of energies of individual vortices and the
interaction potential. In \cite{PeKe11}, a reduced energy for a pair of vortices of the opposite charge
(vortex dipole) was obtained by using the same formal Rayleigh--Ritz method
and analysis of resulting integrals in the limit $\eps \to 0$.

Here we rewrite the result of computations in Lemmas 3 and 4 of \cite{PeKe11}
in the case of a pair of vortices of the same
charge one. We also add the nonlinear dependence of the frequency of vortex precession
on the displacement $a$ from the center of the harmonic potential, which is modeled by
the additional term in the kinetic energy (\ref{kinetic-2}).

Let the two vortices be located
at the distinct points $(x_1,y_1)$ and $(x_2,y_2)$ on the plane
such that $a_1 := (x_1^2+y_1^2)^{1/2}$ and $a_2 := (x_2^2+y_2^2)^{1/2}$ are small, $\eps$ is small,
and $a := ((x_2-x_1)^2+(y_2-y_1)^2)^{1/2}/\eps$ is large. The two-vortex configuration
has kinetic $K$ and potential $P$ energies given at the leading order by
\begin{equation}
\label{kinetic-3}
K(x_1,x_2,y_1,y_2) = \frac{1}{2} \eps \sum_{j=1}^2 (x_j \dot{y}_j - y_j \dot{x}_j) \left[ 1 - \frac{1}{2} (x_j^2 + y_j^2) \right]
\end{equation}
and
\begin{equation}
\label{potential-3}
P(x_1,x_2,y_1,y_2) =  \frac{1}{2} \eps \omega_0(\eps) \sum_{j=1}^2 (x_j^2+y_j^2) + \frac{1}{2} \eps^2 \log\left[ (x_1-x_2)^2 + (y_1-y_2)^2 \right].
\end{equation}

In the reference frame rotating with the angular frequency $\omega$, we can use the polar coordinates
\begin{equation}
\label{rotating-frame-2}
x_j = \xi_j \cos(\omega t) - \eta_j \sin(\omega t), \quad y_j = \xi_j \sin(\omega t) + \eta_j \cos(\omega t), \quad j = 1,2,
\end{equation}
and rewrite the truncated kinetic and potential energies in the form
\begin{eqnarray*}
K(\xi_1,\xi_2,\eta_1,\eta_2) & = & \frac{1}{2} \eps \sum_{j=1}^2 (\xi_j \dot{\eta}_j - \dot{\xi}_j \eta_j)
+ \frac{1}{2} \eps \sum_{j=1}^2 \omega (\xi_j^2+\eta_j^2) \left[ 1 - \frac{1}{2} (\xi_j^2 + \eta_j^2) \right], \\
P(\xi_1,\xi_2,\eta_1,\eta_2) & =  & \frac{1}{2} \eps \omega_0(\eps) \sum_{j=1}^2 (\xi_j^2 + \eta_j^2) +
\frac{1}{2} \eps^2 \log\left[ (\xi_1-\xi_2)^2 + (\eta_1-\eta_2)^2 \right],
\end{eqnarray*}
where the nonlinear correction in $K(\xi_1,\xi_2,\eta_1,\eta_2)$ is dropped to simplify the time evolution of the ODE system.
In the remainder of this section, we obtain the existence and stability results
for two vortices of charge one within the ODE theory.

\subsection{Existence of steadily rotating vortex pairs}

Steadily rotating pairs of vortices are critical points of the action functional
\begin{eqnarray}
\nonumber
E_2(\xi_1,\xi_2,\eta_1,\eta_2) & = & \frac{1}{2} \eps \omega \sum_{j=1}^2 (\xi_j^2+\eta_j^2) \left[ 1 - \frac{1}{2} (\xi_j^2 + \eta_j^2) \right] \\
& \phantom{t} & - \frac{1}{2} \eps \omega_0(\eps) \sum_{j=1}^2 (\xi_j^2 + \eta_j^2) -
\frac{1}{2} \eps^2 \log\left[ (\xi_1-\xi_2)^2 + (\eta_1-\eta_2)^2 \right].
\label{reduced-2}
\end{eqnarray}
We assume that the two vortices are located along the straight line that passes through the center of the harmonic
potential. By using the rotational symmetry of the vortex configuration on the plane,
we select the vortex location at two points
$(\xi_1,\eta_1) = (b_1,0)$ and $(\xi_2,\eta_2) = (-b_2,0)$ for $b_1,b_2 > 0$.
After dividing Euler--Lagrange equations for $E_2(b_1,-b_2,0,0)$ by $\eps$,
we obtain the following system of algebraic equations:
\begin{equation}
\label{pairs-system}
\left\{ \begin{array}{l} \omega b_1 ( 1 - b_1^2) - \omega_0(\eps) b_1 - \eps (b_1+b_2)^{-1} = 0, \\
\omega b_2 ( 1 - b_2^2) - \omega_0(\eps) b_2 - \eps (b_1+b_2)^{-1} = 0, \end{array} \right.
\end{equation}
Subtracting one equation from another, we obtain the constraint
\begin{equation}
\label{pairs}
(b_1-b_2) \left[ \omega - \omega_0(\eps) - \omega (b_1^2+b_1b_2 + b_2^2) \right] = 0.
\end{equation}

\begin{figure}[htb]
\begin{center}
\includegraphics[height=6cm]{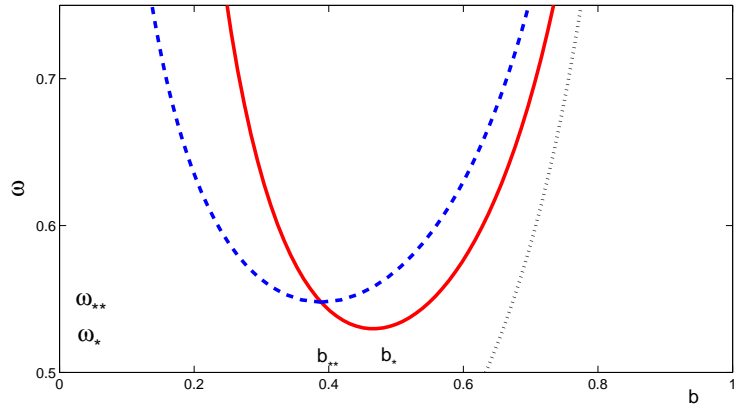}
\end{center}
\caption{A typical example of the bifurcation
  diagram for two vortices of charge one, for $\epsilon=0.05$.
  The symmetric (red, solid) and asymmetric (blue, dashed) pair of vortices
  are shown on the $(b,\omega)$ parameter plane.
The branch of the single vortex displaced from the origin by the distance $b$ is shown by a black dotted line.}
\label{fig-pair}
\end{figure}

The first root in (\ref{pairs}) determines the symmetric vortex pair with $b_1 = b_2 = b$ related to $\omega$ by
\begin{equation}
\label{sym-pair}
\omega(b) = \frac{1}{1-b^2} \left[ \omega_0(\eps) + \frac{\eps}{2b^2} \right].
\end{equation}
The graph of $(0,1) \ni b \mapsto \omega \in \mathbb{R}$ has a global minimum at
the point $(b_*,\omega_*)$, where
\begin{equation}
\label{sym-pair-bif}
2 \omega_* b_*^4 = \eps \quad \Rightarrow \quad \omega_* = \omega_0(\eps) + \frac{\eps}{b_*^2} > \omega_0(\eps).
\end{equation}

The second root in (\ref{pairs}) determines the asymmetric vortex pair with $b_1 \neq b_2$ related to $\omega$ by
the system
\begin{equation}
\label{asym-pair}
\left\{ \begin{array}{l} \omega (1 - b_1^2 - b_1b_2 - b_2^2) = \omega_0(\eps), \\
\omega b_1 b_2 (b_1 + b_2)^2 = \eps, \end{array} \right.
\end{equation}
where the second equation was obtained from system (\ref{pairs-system})
after dividing the first equation by $b_1$, the second equation by $b_2$ and subtracting the result.
The branch of the asymmetric vortex pair bifurcates from the branch
of the symmetric vortex pair at the point $(b_{**},\omega_{**})$, where
\begin{equation}
\label{asym-pair-bif}
4 \omega_{**} b_{**}^4 = \eps \quad \Rightarrow \quad \omega_{**} = \omega_0(\eps) + \frac{3 \eps}{4 b_{**}^2} > \omega_0(\eps).
\end{equation}
Since $(b_*,\omega_*)$ is the only (global) minimum of the graph of $(0,1) \ni b \mapsto \omega \in \mathbb{R}$
and $(b_*,\omega_*)$ is clearly different from $(b_{**},\omega_{**})$, then we have $\omega_{**} > \omega_*$.
Comparing (\ref{sym-pair-bif}) and (\ref{asym-pair-bif}), we obtain $3 b_*^2 > 4 b_{**}^2$ which yields $b_* > b_{**}$.

Figure \ref{fig-pair} represents a typical illustration of branches of the symmetric and asymmetric
vortex pairs on the $(b,\omega)$ parameter plane for $\eps = 0.05$
with the notations used in (\ref{sym-pair-bif}) and (\ref{asym-pair-bif}).
Both branches lie above the branch of a single vortex given by (\ref{omega-law}) with $a = b$.

It should be noted that the symmetry-breaking bifurcation
from the symmetric to the asymmetric vortex pair was identified in the work
of~\cite{Navarro} (see also~\cite{Zamp}). Here, we put this picture in
the context of the stability and variational characterization of the
two-vortex states.

\subsection{Variational characterization of vortex pairs}

Extremal properties of the two critical points of $E_2(\xi_1,\xi_2,\eta_1,\eta_2)$ are studied from
the Hessian matrix $E_2''(b_1,-b_2,0,0)$. This is a block-diagonal matrix in variables $(\xi_1,\xi_2)$ and
$(\eta_1,\eta_2)$ with the two blocks given by
\begin{eqnarray}
\nonumber
L_+ & := & \partial_{\xi_i} \partial_{\xi_j} E_2(b_1,0,-b_2,0) \\
& = & \eps
\left[ \begin{matrix} \omega(1 - 3 b_1^2) - \omega_0(\eps) + \frac{\eps}{(b_1+b_2)^2} & - \frac{\eps}{(b_1+b_2)^2} \\
- \frac{\eps}{(b_1+b_2)^2}  & \omega(1 - 3 b_2^2) - \omega_0(\eps) + \frac{\eps}{(b_1+b_2)^2} \end{matrix} \right]
\label{L-plus-gen}
\end{eqnarray}
and
\begin{eqnarray}
\nonumber
L_- & := & \partial_{\eta_i} \partial_{\eta_j} E_2(b_1,0,-b_2,0) \\
& = & \eps \left[ \begin{matrix}  \omega(1 - b_1^2) - \omega_0(\eps) - \frac{\eps}{(b_1+b_2)^2} & \frac{\eps}{(b_1+b_2)^2}\\
\frac{\eps}{(b_1+b_2)^2} & \omega(1 - b_2^2) - \omega_0(\eps) - \frac{\eps}{(b_1+b_2)^2} \end{matrix} \right].
\label{L-minus-gen}
\end{eqnarray}

Substituting the system (\ref{pairs-system}) into $L_-$ yields a simpler expression
$$
L_- = \frac{\eps^2}{b_1 b_2 (b_1+b_2)^2}
\left[ \begin{matrix}  b_2^2 & b_1 b_2 \\
b_1 b_2 & b_1^2 \end{matrix} \right],
$$
with a simple zero eigenvalue and a simple positive eigenvalue.
The eigenvector $(\xi_1,\xi_2,\eta_1,\eta_2)$ for the zero eigenvalue of $E_2''(b_1,-b_2,0,0)$ is
$R := (0,0,b_1,- b_2)^T$. This eigenvector is related to the rotational invariance of the vortex pair.

Eigenvalues of $L_+$ can be computed with some additional effort. For the symmetric vortex pair with $b_1 = b_2 = b$
and $\omega = \omega(b)$ given by (\ref{sym-pair}), we simplify the entries of $L_+$ as follows
\begin{equation}
\label{L-plus-sym}
L_+ = \eps \left[ \begin{matrix} -2 \omega(b) b^2 + \frac{3 \eps}{4 b^2} & - \frac{\eps}{4 b^2} \\
- \frac{\eps}{4 b^2}  & -2 \omega(b) b^2 + \frac{3 \eps}{4 b^2} \end{matrix} \right].
\end{equation}
The two eigenvalues of $L_+$ are, thus, given by
\begin{equation}
\label{lambda-1-2}
\lambda_1 = -2 \eps \omega(b) b^2 + \frac{\eps^2}{b^2}, \quad \lambda_2 = -2 \eps \omega(b) b^2 + \frac{\eps^2}{2b^2}.
\end{equation}
Increasing $b$ in the interval $(0,1)$, we can detect two bifurcations at $b_{**}$ and $b_*$,
when the eigenvalues pass through the origin. For $b \in (0,b_{**})$, both eigenvalues
of $L_+$ are positive. Hence the critical point $(b,-b,0,0)$ with the smallest displacement
$b$ is a degenerate minimum of $E_2$ with a simple zero eigenvalue (due to $L_-$)
for $\omega > \omega_{**}$. For $b \in (b_{**},b_*)$, we have $\lambda_2 < 0$ and $\lambda_1 > 0$, hence
the critical point $(b,-b,0,0)$ with the smallest displacement $b$
is a saddle point of $E_2$ with one negative ($\lambda_2)$ and one zero (due to $L_-)$
eigenvalues for $\omega \in (\omega_*,\omega_{**})$.
For $b \in (b_{*},1)$, we have $\lambda_1 < 0$ and $\lambda_2 < 0$, hence
the critical point $(b,-b,0,0)$  with the largest displacement $b$
is a saddle point of $E_2$ with two negative $(\lambda_1,\lambda_2)$ and one zero (due to $L_-$)
eigenvalues for $\omega > \omega_*$.

For the asymmetric vortex pair with $b_1 \neq b_2$, we use system (\ref{pairs-system}) and
simplify the entries of $L_+$ as follows
$$
L_+ = \eps \left[ \begin{matrix} -2 \omega b_1^2 + \frac{\eps (2b_1 + b_2)}{b_1 (b_1+b_2)^2} & - \frac{\eps}{(b_1+b_2)^2} \\
- \frac{\eps}{(b_1+b_2)^2}  & -2 \omega b_2^2 + \frac{\eps (b_1 + 2b_2)}{b_2 (b_1+b_2)^2} \end{matrix} \right].
$$
Substituting the second equation of system (\ref{asym-pair}) yields a simpler expression:
\begin{equation}
\label{L-plus-asym}
L_+ = \frac{\eps^2}{b_1 b_2 (b_1 + b_2)^2}
\left[ \begin{matrix} b_2^2 + 2 b_1 b_2 - 2b_1^2 & - b_1 b_2 \\
- b_1 b_2  & b_1^2 + 2 b_1 b_2 - 2b_2^2 \end{matrix} \right],
\end{equation}
with the determinant given by
$$
\det(L_+) = -\frac{2 \eps^4}{b_1^2 b_2^2 (b_1 + b_2)^4} \left[ (b_1^2-b_2^2)^2 + b_1 b_2 (b_1 - b_2)^2 \right].
$$
Since ${\rm det}(L_+) < 0$, the matrix $L_+$ has one negative and one positive eigenvalue. Hence, the
the critical point $(b_1,-b_2,0,0)$  is a saddle point of $E_2$
with one negative (due to $L_+$) and one zero (due to $L_-$)
eigenvalue for all $\omega > \omega_{**}$.

Let us now add the symplectic orthogonality constraint related to the symplectic matrix
\begin{equation}
\label{symplectic-2}
J = \left[ \begin{array}{cccc} 0 & 0 & 1 & 0\\ 0 & 0 & 0 & 1 \\ -1 & 0 & 0 & 0 \\ 0 & -1 & 0 & 0 \end{array} \right],
\end{equation}
which arises in the Hamiltonian system of equations of motion near the vortex pair, see system
(\ref{ham-sys-2}) below. Since $R = (0,0,b_1,- b_2)^T$ is the eigenvector for the zero eigenvalue
of the Hessian matrix $E_2''(b_1,-b_2,0,0)$, the symplectic orthogonality constraint takes the form
\begin{equation}
\label{constraint-2}
\phi := (\xi_1,\xi_2,\eta_1,\eta_2)^T \in \mathbb{R}^4 : \quad
\langle J^{-1} \phi, R \rangle = 0.
\end{equation}
Due to the structure of $J$ and $R$, the constraint simplifies to the equation
\begin{equation}
\label{constraint-2eq}
b_1 \eta_1 - b_2 \eta_2 = 0.
\end{equation}

For the symmetric vortex pair with $b_1 = b_2 = b$, the constraint (\ref{constraint-2eq}) is equivalent to $\eta_1 = \eta_2$.
Projecting $L_+$ in (\ref{L-plus-sym}) to the subspace satisfying this constraint yields
$$
\frac{1}{2} (1,1) L_+ (1,1)^T = -2\eps \omega(b) b^2 + \frac{\eps^2}{2 b^2} = \lambda_2,
$$
where $\lambda_2$ is defined by (\ref{lambda-1-2}).
Since $\lambda_2 > 0$ for $b < b_{**}$ and $\lambda_2 < 0$ for $b > b_{**}$,
the critical point $(b,-b,0,0)$ is a minimizer of $E_2$ for $b < b_{**}$ and
a saddle point of $E_2$ for $b > b_{**}$ under the constraint (\ref{constraint-2}).
No change in the number of negative eigenvalues of $L_+$ constrained by (\ref{constraint-2})
occurs at $b = b_* > b_{**}$, which has only one negative eigenvalue for both $b \in (b_{**},b_*)$ and $b \in (b_*,1)$.

For the asymmetric vortex pair with $b_1 \neq b_2$, projecting $L_+$ in (\ref{L-plus-asym})
to the subspace satisfying the constraint (\ref{constraint-2}) yields
$$
\frac{1}{b_1^2 + b_2^2} (b_2, b_1) L_+ (b_2,b_1)^T
= \frac{\eps^2}{b_1 b_2 (b_1 + b_2)^2 (b_1^2 + b_2^2)} \left[ (b_1^2-b_2^2)^2 + 2 b_1 b_2 (b_1 - b_2)^2 \right] > 0.
$$
Since the operator $L_+$ constrained by (\ref{constraint-2}) is positive,
the critical point $(b_1,-b_2,0,0)$ is a constrained minimizer of $E_2$ under the constraint (\ref{constraint-2}).

\subsection{Stability of vortex pairs}

Stability of the two critical points of $E_2(\xi_1,\xi_2,\eta_1,\eta_2)$ is determined by
equations of motion obtained from the leading-order Lagrangian
\begin{equation}
\label{Lagrangian-2}
L_2(\xi_1,\eta_1,\xi_2,\eta_2) = \frac{1}{2} \eps \sum_{j=1}^2 (\xi_j \dot{\eta}_j - \eta_j \dot{\xi}_j) + E_2(\xi_1,\eta_1,\xi_2,\eta_2).
\end{equation}
After dividing Euler--Lagrange equations by $\eps$, equations of motion take the form
\begin{eqnarray*}
\left\{ \begin{array}{l}
\dot{\eta}_1 + \omega \xi_1 ( 1 - \xi_1^2 - \eta_1^2) - \omega_0(\eps) \xi_1
- \frac{\eps (\xi_1 - \xi_2)}{(\xi_1-\xi_2)^2 + (\eta_1 - \eta_2)^2} = 0,\\
\dot{\eta}_2 + \omega \xi_2 ( 1 - \xi_2^2 - \eta_2^2) - \omega_0(\eps) \xi_2
+ \frac{\eps (\xi_1 - \xi_2)}{(\xi_1-\xi_2)^2 + (\eta_1 - \eta_2)^2} = 0,\\
\dot{\xi}_1 - \omega \eta_1 ( 1 - \xi_1^2 - \eta_1^2) + \omega_0(\eps) \eta_1
+ \frac{\eps (\eta_1 - \eta_2)}{(\xi_1-\xi_2)^2 + (\eta_1 - \eta_2)^2} = 0,\\
\dot{\xi}_2 - \omega \eta_2 ( 1 - \xi_2^2 - \eta_2^2) + \omega_0(\eps) \eta_2
- \frac{\eps (\eta_1 - \eta_2)}{(\xi_1-\xi_2)^2 + (\eta_1 - \eta_2)^2} = 0, \end{array} \right.
\end{eqnarray*}
which can be written as the Hamiltonian system
\begin{equation}
\label{ham-sys-2}
\frac{d}{d t} \left( \begin{array}{c} \xi_1 \\ \xi_2 \\ \eta_1 \\ \eta_2 \end{array} \right) =
J \left( \begin{array}{c} \frac{\partial E_2}{\partial \xi_1} \\ \frac{\partial E_2}{\partial \xi_2} \\
\frac{\partial E_2}{\partial \eta_1} \\ \frac{\partial E_2}{\partial \eta_2} \end{array} \right),
\end{equation}
where $E_2$ in (\ref{reduced-2}) serves as the Hamiltonian function and $J$ is defined by (\ref{symplectic-2}).

Linearizing equations of motion at
the critical point $(\xi_1,\xi_2,\eta_1,\eta_2) = (b_1,-b_2,0,0)$ with
$$
\xi_1 = b_1 + \hat{\xi}_1 e^{\lambda t}, \quad
\xi_2 = -b_2 + \hat{\xi}_2 e^{\lambda t}, \quad
\eta_1 = \hat{\eta}_1 e^{\lambda t}, \quad
\eta_2 = \hat{\eta}_2 e^{\lambda t}
$$
yields the spectral stability problem
\begin{equation}
\label{spectral-stab}
L_+ \hat{\xi} = - \lambda \hat{\eta}, \quad L_- \hat{\eta} = \lambda \hat{\xi},
\end{equation}
where $\hat{\xi} = (\hat{\xi}_1,\hat{\xi}_2)^T$, $\hat{\eta} = (\hat{\eta}_1,\hat{\eta}_2)^T$,
whereas $L_+$ and $L_-$ are given by (\ref{L-plus-gen}) and (\ref{L-minus-gen}).

For the symmetric vortex pair with $b_1 = b_2 = b$, the spectral stability problem
(\ref{spectral-stab}) can be block-diagonalized into two decoupled problems:
\begin{eqnarray}
\label{block-1}
\left\{ \begin{array}{l}
\left[ - 2 \omega(b) b^2 + \frac{\eps}{2 b^2} \right] (\hat{\xi}_1 + \hat{\xi}_2) = - \lambda (\hat{\eta}_1 + \hat{\eta}_2),\\
\frac{\eps}{2 b^2} (\hat{\eta}_1 + \hat{\eta}_2) = \lambda (\hat{\xi}_1 + \hat{\xi}_2) \end{array} \right.
\end{eqnarray}
and
\begin{eqnarray}
\label{block-2}
\left\{ \begin{array}{l}
\left[ - 2 \omega(b) b^2 + \frac{\eps}{b^2} \right] (\hat{\xi}_1 - \hat{\xi}_2) = - \lambda (\hat{\eta}_1 - \hat{\eta}_2),\\
0 = \lambda (\hat{\xi}_1 - \hat{\xi}_2). \end{array} \right.
\end{eqnarray}
The second block (\ref{block-2}) yields a double zero eigenvalue with a non-diagonal Jordan block. The double
zero eigenvalue is related to the rotational invariance of the symmetric vortex pair.
The first block (\ref{block-1}) yields a symmetric pair of eigenvalues from the characteristic equation
$$
\lambda^2 = \frac{\eps}{2 b^2} \left[ 2 \omega(b) b^2 - \frac{\eps}{2 b^2} \right] = -\frac{1}{2b^2} \lambda_2,
$$
where $\lambda_2$ is defined by (\ref{lambda-1-2}).
Since $\lambda_2 > 0$ for $b < b_{**}$ and $\lambda_2 < 0$ for $b > b_{**}$,
we have $\lambda^2 < 0$ for $b < b_{**}$ and $\lambda^2 > 0$ for $b > b_{**}$.
Hence the symmetric vortex pair is stable with $b < b_{**}$ and unstable for $b > b_{**}$ with exactly
one pair of real eigenvalues. This agrees with the variational characterization of the
critical point $(b,-b,0,0)$, which is a minimizer of $E_2$ for $b < b_{**}$ and a constrained saddle point of
$E_2$ for $b > b_{**}$ under the constraint (\ref{constraint-2}).

For the asymmetric vortex pair with $b_1 \neq b_2$, the spectral stability problem
(\ref{spectral-stab}) has again a double zero eigenvalue with a non-diagonal Jordan block,
thanks to the rotational invariance of the vortex pair. It remains to find the other pair
of eigenvalues $\lambda$. To eliminate the translational invariance, let us assume that
$b_2 \hat{\eta}_1 + b_1 \hat{\eta}_2 \neq 0$, then $(\hat{\xi},\hat{\eta}) \nparallel R = (0,0,b_1,-b_2)^T$.
If this is the case, we find from the spectral problem (\ref{spectral-stab}) that
$$
\lambda b_1 \hat{\xi}_1 = \lambda b_2 \hat{\xi}_2 = \frac{\eps}{(b_1+b_2)^2} (b_2 \hat{\eta}_1 + b_1 \hat{\eta}_2).
$$
after which the symmetric pair of eigenvalues is determined by the characteristic equation
\begin{eqnarray*}
\lambda^2 = -\frac{\eps^2}{b_1^2 b_2^2 (b_1+b_2)^4} \left[ (b_1^2-b_2^2)^2 + 2 b_1 b_2 (b_1 - b_2)^2 \right].
\end{eqnarray*}
Since $\lambda^2 < 0$, the asymmetric vortex pair is stable for all $\omega > \omega_{**}$.
This agrees with the variational characterization of the critical point $(b_1,-b_2,0,0)$,
which is a constrained minimizer of $E_2$ under the constraint (\ref{constraint-2}).

\section{Numerical results for the Gross--Pitaevskii equation}

To complement the ODE theory, we present
direct numerical simulations of the PDE model (\ref{GP}) for a small
value of $\eps$. In particular, we set $\eps=0.05$.

The two-vortex solutions are identified in a co-rotating
frame with frequency $\omega$ (in which case the solutions are stationary
and can be obtained by a Newton-type iteration).
Both the symmetric and the asymmetric branches of the two-vortex solutions
are obtained in this way. For the former, in line with the theoretical prediction
on Fig. \ref{fig-pair}, a bifurcation point is identified at $\omega_* \approx 0.587$,
the symmetric two-vortex solutions can only be obtained
for $\omega> \omega_{*}$. The resulting solutions can be found both
with $b>b_{*}$ and with $b<b_{*}$. The numerical value $b_* \approx 0.522$
from the PDE model is close to the predicted value $b_*^{(th)} \approx 0.490$
from the ODE theory. For the branch of symmetric two-vortex solutions
with $b<b_*$, a second bifurcation point is identified at $\omega_{**} \approx 0.693$
and the pair of asymmetric two-vortex  solutions is obtained for $\omega > \omega_{**}$.
The numerical value $b_{**} \approx 0.352$ is again close to the predicted value
$b_{**}^{(th)} \approx 0.408$.

Although the ODE theory captures fully the bifurcation diagram
of the PDE model, there are some quantitative differences in the
bifurcation points. The differences exist because the ODE theory is
valid in the semi-classical limit $\eps \to 0$, whereas
the PDE model is studied at a fixed finite $\eps = 0.05$.

\begin{figure}[htb]
\begin{center}
  \includegraphics[height=6cm,width=8cm]{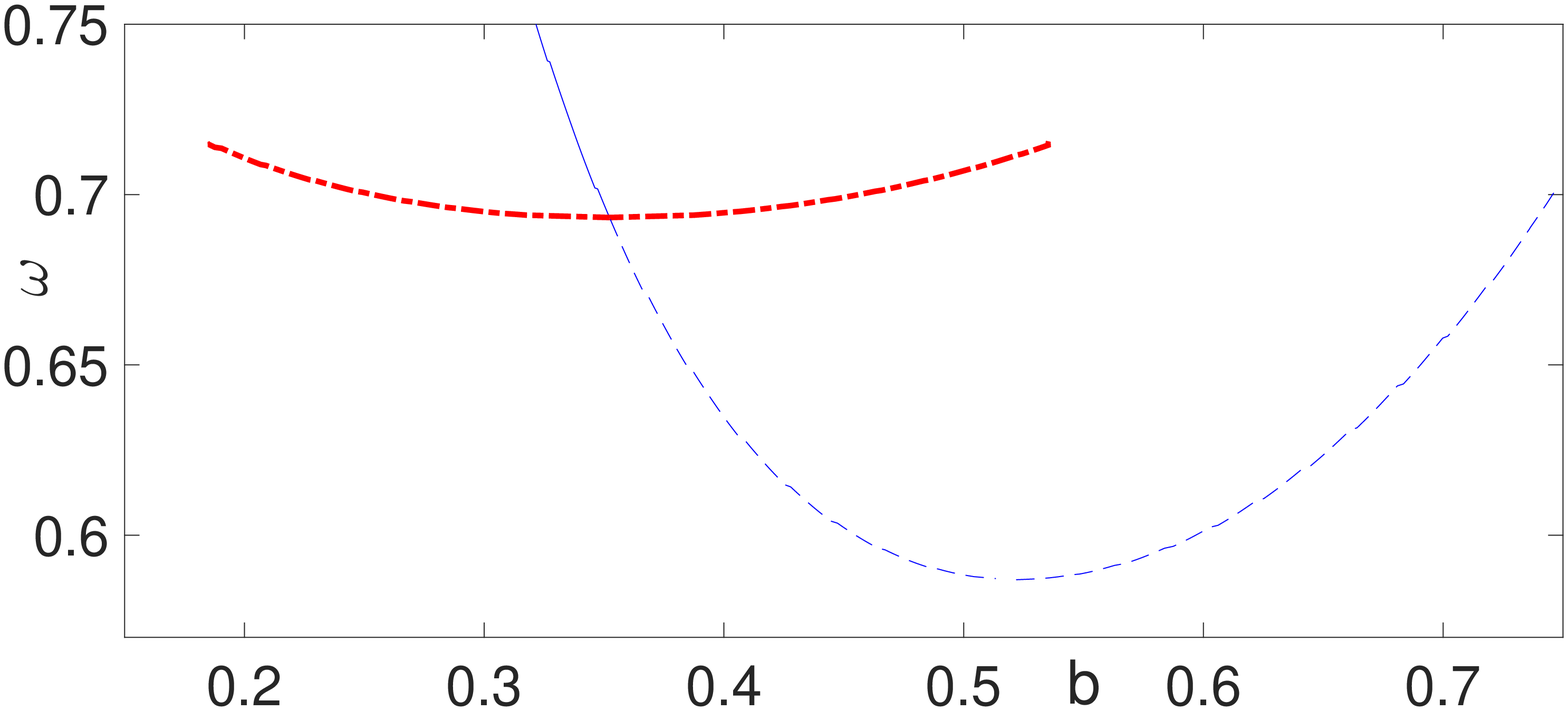}
  \includegraphics[height=6cm,width=8cm]{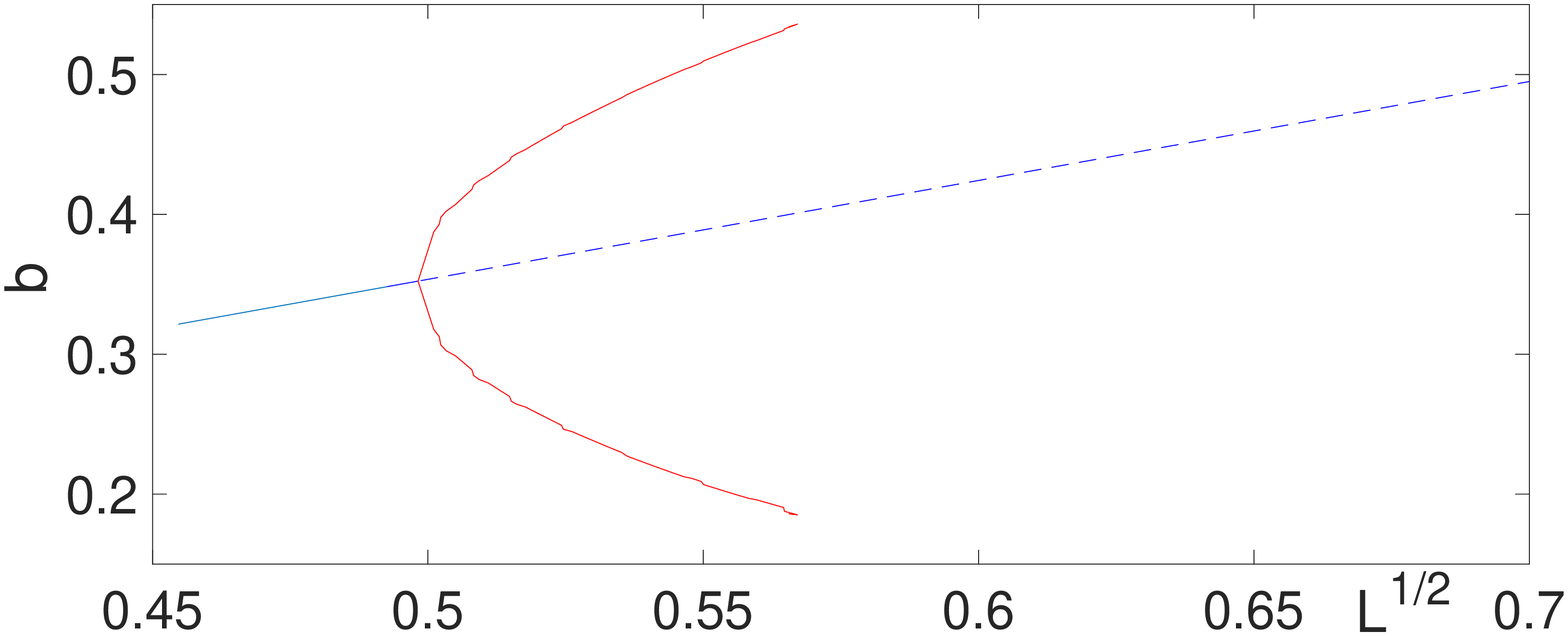}
\end{center}
\caption{Bifurcation diagram of the symmetric and asymmetric vortex pairs for $\epsilon=0.05$. The left panel
  shows the diagram in the vortex position-rotation frequency
  variables $(b,\omega)$. The solid line corresponds to the
  spectrally stable symmetric vortex pair, the dash-dotted
  one corresponds to the unstable symmetric vortex pair, while the thick dash dotted
  branch corresponds to the stable asymmetric vortex pair. The right panel
  shows the bifurcation diagram in the variables $(b,L)$ with $L = b_1^2 + b_2^2$.}
\label{dpfig1}
\end{figure}

\begin{figure}[tb]
\begin{center}
\includegraphics[height=6cm]{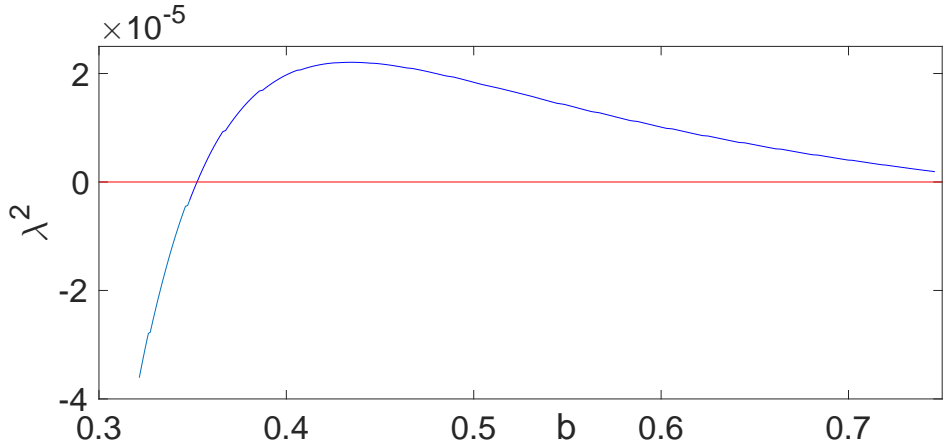}
\end{center}
\caption{Squared eigenvalues of the spectral stability problem for the symmetric vortex pair.
  The unstable eigenvalue with $\lambda^2 > 0$ exists for $b > b_{**}$ in agreement with the ODE theory.}
\label{dpfig2}
\end{figure}

\begin{figure}
\begin{center}
\includegraphics[height=6cm,width=5cm]{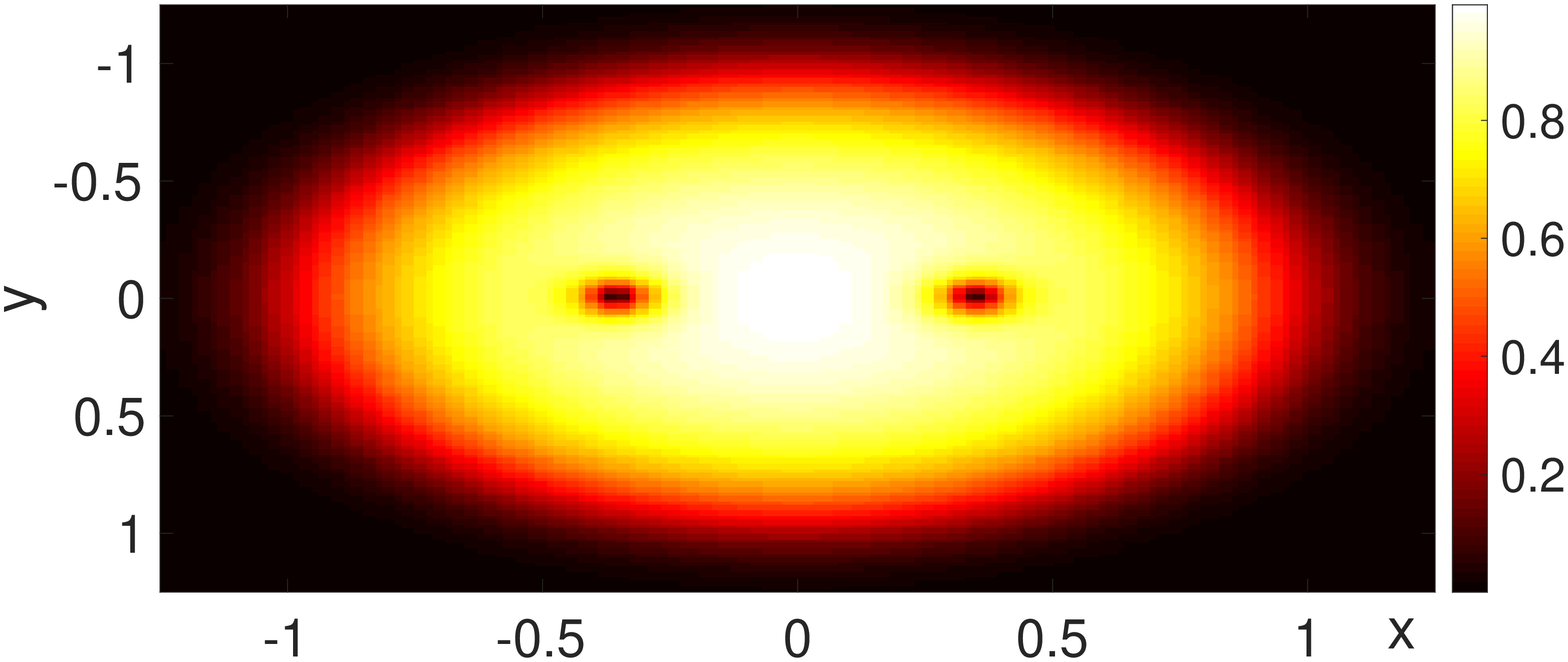}
\includegraphics[height=6cm,width=5cm]{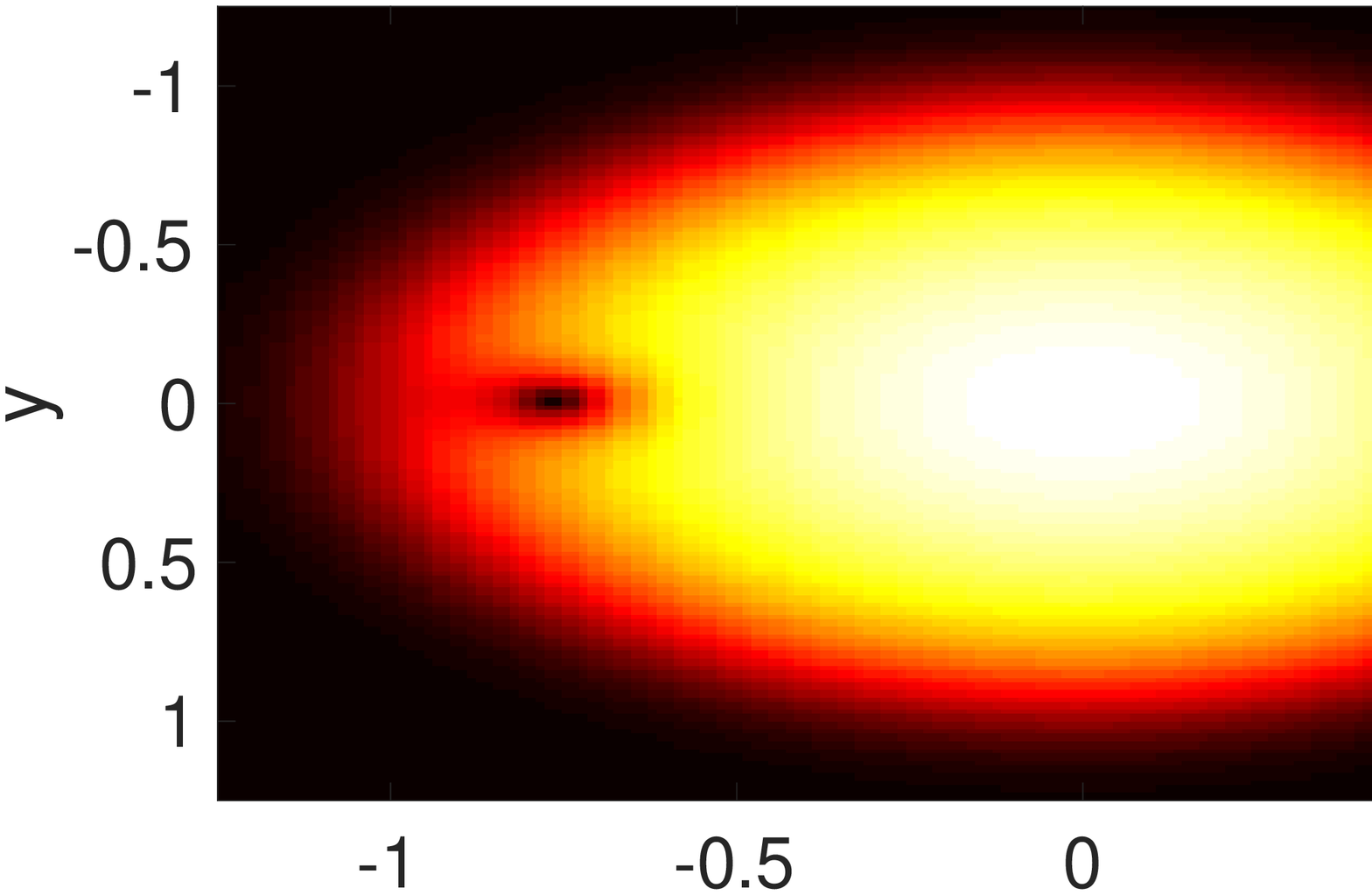}
\includegraphics[height=6cm,width=5cm]{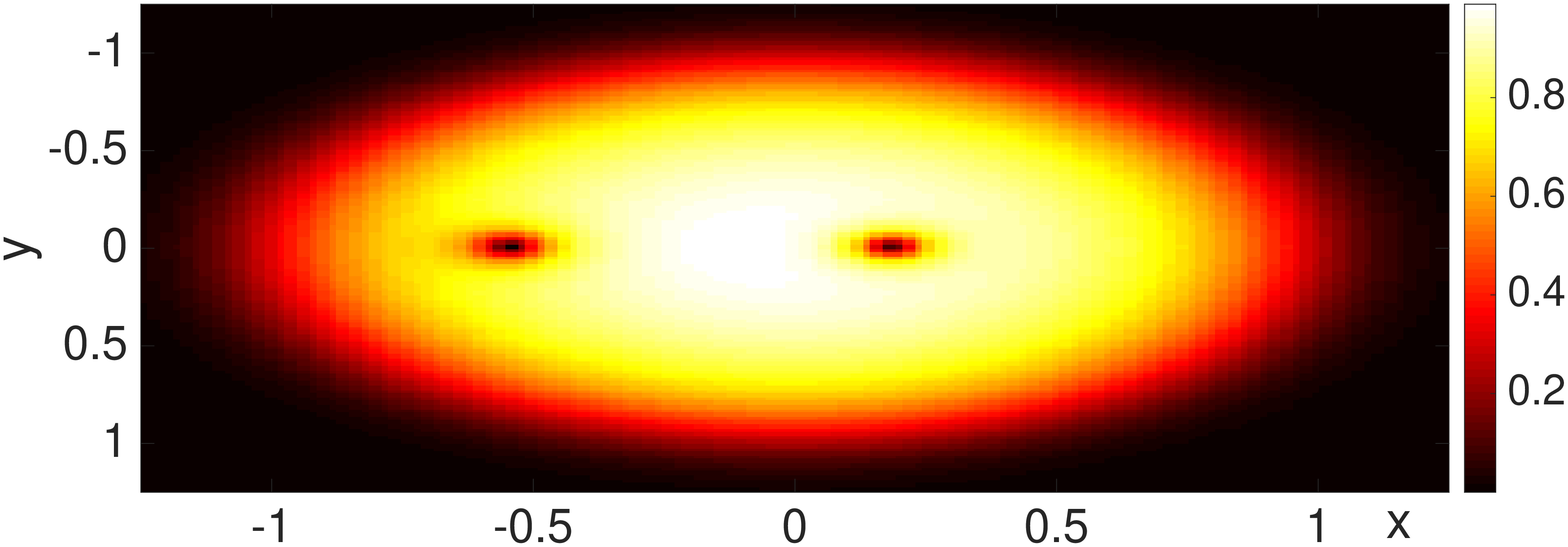}
\end{center}
\caption{Left and middle panels: two examples of the symmetric vortex pair for the same value of $\omega=0.7$.
Right panel: an example of the asymmetric vortex pair for $\omega=0.715$.}
\label{dpfig3}
\end{figure}

The different branches of the bifurcation diagram in
the $(b,\omega)$ variables are shown in the
left panel of Fig.~\ref{dpfig1}, in agreement with Fig. \ref{fig-pair}.
The right panel of Fig. \ref{dpfig1} shows the same diagram in the $(b,L)$ variables,
where $L = b_1^2 + b_2^2$ to showcase
the supercritical character of the relevant pitchfork bifurcation,
in agreement with the diagrams used in~\cite{Zamp}.

Fig.~\ref{dpfig2} shows the squared eigenvalue of the spectral stability
problem for the symmetric two-vortex solution. The dependence illustrates
the destabilizing nature of the bifurcation at $\omega = \omega_{**}$ but not at
$\omega = \omega_*$. Indeed, $\lambda^2 < 0$ for $b < b_{**}$ but $\lambda^2 > 0$
for both $b \in (b_{**},b_*)$ and $b \in (b_*,1)$, hence the symmetric
two-vortex solution with $b > b_{**}$ is linearly unstable.

To manifest some typical profiles of the relevant configurations,
in Fig.~\ref{dpfig3}, we show two examples of the symmetric configuration
for the same value of $\omega=0.7$. This serves as a partial illustration
of the ``folded'' nature of the relevant branch of solutions, such that
for each value of $\omega>\omega_{*}$, there exists a pair of symmetric two-vortex
solutions (each of which is invariant under angular rotations).
One of these (left panel) corresponds
to the smaller-than-critical distance, while the other one (middle panel) corresponds to the
larger-than-critical distance. In the latter case, the vortices are nearly at the edges of the cloud.
The right panel illustrates an example of the asymmetric two-vortex solution
for a value of $\omega=0.715$.

\section{Conclusion}

We have revisited the existence and stability of two-vortex configurations
in the context of rotating Bose-Einstein condensates.
As a preamble to the ODE theory, we have discussed
the existence and stability properties of a single
vortex of charge one: the symmetric vortex is located at the center of the trap and
the asymmetric vortex is located at the periphery of the trap.
We showed that the latter bifurcates at $\omega = \omega_0(\eps)$,
where $\omega_0(\eps)$ is the linear eigenfrequency of precession of
a single vortex near the center of the trap in the absence of rotation.
The symmetric vortex is an energy minimizer for $\omega > \omega_0(\eps)$,
whereas the asymmetric vortex is a constrained energy minimizer
under the constraint eliminating rotational invariance.

We have also considered the relevant two-vortex configurations, when both vortices have
the same charge one. In this context, the symmetric vortex pair bifurcates
at $\omega = \omega_*$ via the saddle-node bifurcation of two different vortex pairs,
whereas the asymmetric vortex pair bifurcates at $\omega = \omega_{**}$ via the supercritical
pitchfork bifurcation. The symmetric vortex pairs exist for $\omega > \omega_*$
and the two distinct solutions have either smaller-than-critical or larger-than-critical distance
from the center of the trap. The asymmetric vortex pairs exist for $\omega > \omega_{**}$ and bifurcate
from the symmetric vortex pair with the smaller-than-critical distance from the center of the trap.
The two vortices in the asymmetric vortex pair are located at unequal distances from the trap center.
We showed that the symmetric vortex pair with the smaller-than-critical distance
is an energy minimizer for $\omega > \omega_{**}$, whereas the asymmetric vortex pair
is a constrained energy minimizer under the constraint eliminating rotational invariance.
We also showed that all other symmetric vortex pairs are unstable as they are
saddle points of the energy even under the same constraint.

The ODE theory is compared with the full numerical approximations of the PDE model
and a good correspondence is established for $\eps = 0.05$.

\appendix

\section{Derivation of the asymptotic expansion (\ref{kinetic-2})}

The kinetic energy $K(x_0,y_0)$ of a single vortex given by the asymptotic expansion
(\ref{kinetic-1}) is determined in \cite{PeKe11} from the expression
$$
K = \frac{i \eps}{4\pi} \int_{\mathbb{R}^2} \eta_{\eps}^2 ( v \bar{v}_t - \bar{v} v_t) dx,
$$
where $\eta_{\eps}$ is the positive real radially-symmetric ground state and
$v$ is represented by the free vortex solution of the defocusing
nonlinear Schr\"{o}dinger equation placed at the point $(x_0,y_0)$. After substitution and separation
of variables, the following expansion was obtained in the proof of Lemma 1 in \cite{PeKe11}:
$$
K = - \dot{x}_0 K_x - \dot{y}_0 K_y,
$$
where
\begin{eqnarray*}
K_x & = & - \frac{\eps^2}{2 \pi} \int_{\mathbb{R}^2} \eta_{\eps}^2(|x|) \frac{Y}{R^2} dX dY + \mathcal{O}(\eps^2 |y_0|), \\
K_y & = & \frac{\eps^2}{2 \pi} \int_{\mathbb{R}^2} \eta_{\eps}^2(|x|) \frac{X}{R^2} dX dY + \mathcal{O}(\eps^2 |x_0|),
\end{eqnarray*}
with $x = x_0 + \eps X$, $y = y_0 + \eps Y$, and $R = (X^2+Y^2)^{1/2}$.

Here we will extend the asymptotic expansion (\ref{kinetic-1}) and will include
the higher-order behavior of $K(x_0,y_0)$ in $(x_0,y_0)$ at the leading order in $\eps$.
By the symmetry of integrals, it is sufficient to analyze the leading order
in the expression for $K_x$ as a function of $y_0$ for $x_0 = 0$.
Therefore, we define
$$
J(y_0) := - \frac{\eps^2}{2 \pi} \int_{\mathbb{R}^2} \eta_{\eps}^2(r) \biggr|_{r = \sqrt{\eps^2 X^2 + (y_0+\eps Y)^2}} \frac{Y}{R^2} dX dY.
$$
Since $J$ is smooth and $J(-y_0) = -J(y_0)$, we have $J(0) = J''(0) = J^{(4)}(0) = 0$.
The first odd derivatives of $J$ can be computed with the chain rule:
\begin{eqnarray*}
J'(0) & = & - \frac{\eps^2}{2 \pi} \int_{\mathbb{R}^2} \partial_r \eta_{\eps}^2(r) |_{r = \eps R} \frac{Y^2}{R^3} dX dY \\
& = & - \frac{\eps^2}{2 \pi} \left[ \int_{0}^{\infty} \partial_r \eta_{\eps}^2(r) |_{r = \eps R} d R \right]
\left[ \int_0^{2\pi} \sin^2\theta d \theta \right] \\
& = & - \frac{\eps}{2} \int_{0}^{\infty} \partial_r \eta_{\eps}^2(r) dr \\
& = & \frac{\eps}{2} \eta_{\eps}(0)^2
\end{eqnarray*}
and
\begin{eqnarray*}
J'''(0) & = & - \frac{\eps^2}{2 \pi} \int_{\mathbb{R}^2}
\left[ \partial^3_r \eta_{\eps}^2(r) |_{r = \eps R} \frac{Y^4}{R^5}
+ 3 \partial^2_r \eta_{\eps}^2(r) |_{r = \eps R} \frac{X^2 Y^2}{\eps R^6}
- 3 \partial_r \eta_{\eps}^2(r) |_{r = \eps R} \frac{X^2 Y^2}{\eps^2 R^7} \right] dX dY  \\
& = & - \frac{3 \eps}{8} \int_{0}^{\infty} \left[ \partial^3_r \eta_{\eps}^2(r)
+ \frac{1}{r} \partial^2_r \eta_{\eps}^2(r) - \frac{1}{r^2}\partial_r \eta_{\eps}^2(r) \right] dr \\
& = & \frac{3 \eps}{8} \lim_{r \to 0} \left[ \partial^2_r \eta_{\eps}(r)^2 + \frac{1}{r} \partial_r \eta_{\eps}^2(r) \right].
\end{eqnarray*}
Let us recall the approximation of $\eta_{\eps}$ with the Thomas--Fermi limit
$$
\eta_0(x) := \lim_{\eps \to 0} \eta_{\eps}(x) = \left\{ \begin{array}{l} (1 - |x|^2)^{1/2}, \quad |x| \leq 1, \\
0, \qquad \qquad |x| > 1, \end{array} \right.
$$
which has been justified in \cite{GalloPel,IM1}. By Proposition 2.1 in \cite{IM1},
for any compact subset $K$ inside the unit disk, there is $C_K > 0$ such that
$$
\| \eta_{\eps} - \eta_0 \|_{C^2(K)} \leq C_K \eps^2.
$$
By using this bound, we compute $J'(0)$ and $J'''(0)$ as $\eps \to 0$:
\begin{eqnarray*}
J'(0) = \frac{\eps}{2} \left[ 1 + \mathcal{O}(\eps^2) \right] \quad \mbox{\rm and} \quad
J'''(0) = -\frac{3 \eps}{2} \left[ 1 + \mathcal{O}(\eps^2) \right],
\end{eqnarray*}
from which we conclude that
$$
J(y_0) = \frac{1}{2} \eps y_0 \left[ 1 - \frac{1}{2} y_0^2 + \mathcal{O}(\eps + y_0^4) \right].
$$
By the symmetry of $K_x$ and similar computations for $K_y$, we obtain the expansion (\ref{kinetic-2}).


\begin{thebibliography}{99}
\bibitem{Bet1} F. Bethuel, R.L. Jerrard, and D. Smets, \emph{On the NLS dynamics for infinite energy vortex configurations
on the plane}, Rev. Mat. Iberoam. {\bf 24} (2008), 671--702.

\bibitem{Bet2} F. Bethuel and J-C. Saut, \emph{Travelling waves for the Gross--Pitaevskii equation},
Ann. Inst. H. Poincare Phys. Theor. {\bf 70} (1999), 147--238.

\bibitem{Bizon} A. Biasi, P. Bizon, B. Craps, and O. Evnin,
\emph{Exact LLL Solutions for BEC Vortex Precession}, arXiv:1705.00867 (2017)

\bibitem{theo2} R. Carretero-Gonz\'{a}lez, P.G. Kevrekidis, and T.
Kolokolnikov, \emph{Vortex nucleation in a dissipative variant of the nonlinear
Schr\"{o}dinger equation under rotation}, Phys. D \textbf{317} (2016), 1--14.

\bibitem{Farell} E.G. Charalampidis, P.G. Kevrekidis, and P.E. Farrell, \emph{Computing stationary
solutions of the 2D Gross--Pitaevskii equation with deflated continuation}, arXiv:1612.08145 (2017)

\bibitem{Chiron} D. Chiron and C. Scheid, \emph{Multiple branches of travelling waves for the
Gross--Pitaevskii equation}, hal-01525255 (2017).

\bibitem{CoGa15} A. Contreras and C. Garc\'{\i}a-Azpeitia. \emph{Global
Bifurcation of Vortices and Dipoles in Bose-Einstein Condensates}, C. R. Math.
Acad. Sci. Paris \textbf{354} (2016), 265--269.

\bibitem{Castin} Y. Castin and R. Dum, \emph{Bose--Einstein condensates with
vortices in rotating traps}, European Phys. J. D \textbf{7} (1999), 399--412.

\bibitem{Bartek} I. Danaila and B. Protas, \emph{Computation of ground states of the
Gross-Pitaevskii functional via Riemannian optimization}, arXiv: 1703.07693 (2017).

\bibitem{review} A.L. Fetter, ``Rotating trapped Bose-Einstein condensates'',
Rev. Mod. Phys. \textbf{81} (2009), 647--691.

\bibitem{GalloPel} C. Gallo and D. Pelinovsky, ``On the Thomas-Fermi ground state
in a harmonic potential", Asymptotic Analysis {\bf 73} (2011), 53--96.

\bibitem{GarciaPel} C.~Garc\'{\i}a-Azpeitia and D.E. Pelinovsky,
\emph{Bifurcations of multi-vortex configurations in rotating Bose-Einstein condensates},
arXiv:1701.01494 (2017).

\bibitem{GKC} R.H. Goodman, P.G. Kevrekidis, and R. Carretero-Gonz\'{a}lez,
\emph{Dynamics of Vortex Dipoles in Anisotropic Bose-Einstein Condensates}
SIAM J. Appl. Dyn. Syst. {\bf 14} (2014), 699--729.

\bibitem {Hani} P. Germain, Z. Hani, and L. Thomann, \emph{On the continuous
resonant equation for NLS. I. Deterministic analysis}, J. Math. Pures Appl.
\textbf{105} (2016), 131--163.

\bibitem{IM1} R. Ignat and V. Millot, \emph{The critical velocity for vortex
existence in a two-dimensional rotating Bose--Einstein condensate}, J. Funct.
Anal. \textbf{233} (2006), 260--306.

\bibitem{IM2} R. Ignat and V. Millot, \emph{Energy expansion and vortex location
for a two-dimensional rotating Bose--Einstein condensate}, Rev. Math. Phys.
\textbf{18} (2006), 119--162.

\bibitem{Kap1} T. Kapitula, P.G. Kevrekidis, and R. Carretero--Gonz\'alez,
\emph{Rotating matter waves in Bose--Einstein condensates}, Physica D \textbf{233}
(2007), 112--137.

\bibitem{siambook} P. G. Kevrekidis, D. J. Frantzeskakis, and R. Carretero-
Gonz{\'a}lez, {\it The Defocusing Nonlinear Schr{\"o}dinger Equation}, SIAM (Philadelphia, 2015).

\bibitem{Kollar} R. Kollar and R.L. Pego, \emph{Spectral stability of vortices in
two-dimensional Bose--Einstein condensates via the Evans function and Krein
signature}, Appl. Math. Res. eXpress \textbf{2012} (2012), 1--46.

\bibitem{theo1} T. Kolokolnikov, P.G. Kevrekidis, and R.
Carretero--Gonz\'alez, \emph{A tale of two distributions: from few to many vortices
in quasi-two-dimensional Bose-Einstein condensates}, Proc. R. Soc. Lond. Ser. A
Math. Phys. Eng. Sci. \textbf{470} (2014), 20140048 (18 pp).

\bibitem{21} P. Kuopanportti, J. A. M. Huhtam\"{a}ki, and M. M\"{o}tt\"{o}nen,
\emph{Size and dynamics of vortex dipoles in dilute Bose-Einstein
condensates,} Phys. Rev. A {\bf 83} (2011), 011603.

\bibitem{34} S. Middelkamp, P. J. Torres, P. G. Kevrekidis, D. J.
Frantzeskakis, R. Carretero-Gonzalez, P. Schmelcher, D. V. Freilich, and D. S.
Hall, \emph{Guiding-center dynamics of vortex dipoles in Bose-Einstein
condensates, } Phys. Rev. A {\bf 84} (2011), 011605.

\bibitem{38} M. M\"{o}tt\"{o}nen, S. M. M. Virtanen, T. Isoshima, and M. M.
Salomaa, \emph{Stationary vortex clusters in nonrotating Bose-Einstein
condensates,} Phys. Rev. A {\bf 71} (2005), 033626.

\bibitem{simula} A.V. Murray, A.J. Grosjek, P. Kuopanportti,
  and T. Simula,
\emph{Hamiltonian dynamics of two same-sign point vortices},
  Phys. Rev. A {\bf 93}, 033649 (2016).

\bibitem{Navarro} R. Navarro, R. Carretero--Gonz\'alez, P.J. Torres, P.G.
Kevrekidis, D.J. Frantzeskakis, M.W. Ray, E. Altuntas, and D.S. Hall,
\emph{Dynamics of a few corotating vortices in Bose--Einstein condensates},
Phys. Rev. Lett. \textbf{110} (2013), 225301.

\bibitem{pitas} L. P. Pitaevskii and S. Stringari,
  {\it Bose-Einstein Condensation}, Oxford University Press (Oxford, 2003).

\bibitem{PeKe11} D. Pelinovsky and P.G. Kevrekidis, \emph{Variational
approximations of trapped vortices in the large-density limit}, Nonlinearity
{\bf 24} (2011), 1271--1289.

\bibitem{PeKe13} D. Pelinovsky and P.G. Kevrekidis, \emph{Bifurcations of
Asymmetric Vortices in Symmetric Harmonic Traps}, Applied Mathematics Research
eXpress {\bf 2013} (2013), 127--164.

\bibitem{Seiringer} R. Seiringer, \emph{Gross-Pitaevskii theory of the rotating
Bose gas}, Commun. Math. Phys. \textbf{229} (2002), 491--509.

\bibitem{57} P. J. Torres, P. G. Kevrekidis, D. J. Frantzeskakis, R.
Carretero-Gonzalez, P. Schmelcher, and D. S. Hall, \emph{Dynamics of vortex
dipoles in confined Bose-Einstein condensates,} Phys. Lett. A {\bf 375} (2011), 3044--3050.

\bibitem{bisset} W. Wang, R.N. Bisset, C. Ticknor,
  R. Carretero-Gonz{\'a}lez, D. J. Frantzeskakis, L. A. Collins, and
  P. G. Kevrekidis,
  \emph{Single and multiple vortex rings in three-dimensional Bose-Einstein condensates: Existence, stability, and dynamics},
Phys. Rev. A {\bf 95} (2017), 043638.

\bibitem{theo} S. Xie, P.G. Kevrekidis, and Th. Kolokolnikov,
\emph{Multi-vortex crystal lattices in Bose-Einstein condensates with a
rotating trap}, preprint (August, 2017).

\bibitem{Zamp} A.V. Zampetaki, R. Carretero-Gonz{\'a}lez, P.G. Kevrekidis,
  F.K. Diakonos, and D.J. Frantzeskakis,
  \emph{Exploring rigidly rotating vortex configurations and their bifurcations in atomic Bose-Einstein condensates},
  Phys. Rev. E {\bf 88} (2013), 042914.


\end{thebibliography}
\end{document}